\documentclass[12pt,preprint]{aastex}

\slugcomment{Submitted to The Astrophysical Journal}

\shorttitle{DR21(OH)}
\shortauthors{Araya, et al.}

\def \Lo {$\,$L$_{\odot}$}
\def \Mo {$\,$M$_{\odot}$}
\def \kms {$\,$km s$^{-1}$}
\def \mjyb {$\,$mJy$\,$beam$^{-1}$}
\def \h{$^{\rm h}$}
\def \m{$^{\rm m}$}
\def\sec{\hbox{$.\!\!^{\rm s}$}}
\def \ad {$\arcdeg$}
\def \am {$\arcmin$}
\def \ads {\hbox{$.\!\!\arcsec$}}

\begin{document}

\title{Radio Continuum and Methanol Observations of DR21(OH)}

\author{Esteban D. Araya\footnote{E. D. Araya is a Jansky Fellow
of the National Radio Astronomy Observatory.}}
\affil{National Radio Astronomy Observatory, P.O. Box 0, 
Socorro, NM 87801; and
Department of Physics and Astronomy, MSC07 4220, 
University of New Mexico, Albuquerque, NM 87131.}

\author{Stan Kurtz}
\affil{Centro de Radioastronom\'{\i}a y Astrof\'{\i}sica, 
Universidad Aut\'onoma de M\'exico,
Apdo. Postal 3-72, 58089, Morelia, Michoac\'an, Mexico.}

\author{Peter Hofner}
\affil{New Mexico Institute of Mining and Technology, 
Physics Department, 801 Leroy Place, Socorro, NM 87801;
and National Radio Astronomy Observatory, P.O. 
Box 0, Socorro, NM 87801.}

\author{Hendrik Linz}
\affil{Max--Planck--Institut f\"ur Astronomie, K\"onigstuhl 17,
D--69117 Heidelberg, Germany.}

\begin{abstract}

We report high sensitivity sub-arcsecond angular resolution
observations of the massive star forming region DR21(OH)
at 3.6, 1.3, and 0.7$\,$cm obtained with the Very Large
Array. In addition, we conducted 
observations of CH$_3$OH 44$\,$GHz masers.
We detected more than 30 new maser components in the 
DR21(OH) region. Most of the masers appear to trace a 
sequence of bow-shocks in a bipolar outflow. 
The cm continuum observations reveal a cluster of radio
sources; the strongest emission is found toward the molecular
core MM1. The radio sources in MM1 are located
about $5^{\prime\prime}$ north of the symmetry center of
the CH$_3$OH outflow, and therefore, they are unlikely 
to be associated with the outflow. Instead, the driving source of the
outflow is likely located in the MM2 core.
Although based on circumstantial evidence, 
the radio continuum from MM1 appears to trace 
free-free emission from shock-ionized gas in a jet.
The orientation of the putative jet in MM1 is approximately 
parallel to the CH$_3$OH outflow and almost perpendicular to 
the large scale molecular filament that connects DR21 and DR21(OH).
This suggests that the (accretion) disks associated with the 
outflows/jets in the DR21 -- DR21(OH) region have
symmetry axes mostly perpendicular to the filament.

\end{abstract}


\section{Introduction}

Located about 3\arcmin~ north of DR21, the DR21(OH) region,
also known as W75S, is a site of dense molecular gas within the
Cygnus-X complex. Both DR21 and DR21(OH) have been extensively
studied in the infrared, sub-millimeter, and millimeter bands.
Both are regions of massive star formation, but DR21 is 
in a more evolved state (as evident from the presence of ultracompact
H{~\small II} regions) while DR21(OH) is in an earlier phase, 
in which the massive (proto)stars have not yet substantially ionized
the surrounding molecular gas.

The DR21(OH) region has four principal condensations, i.e., 
DR21(OH)-N, M, W, and S (Mangum et al. 1992). 
DR21(OH)-M contains two main clumps (MM1 and MM2) 
that are warm, 60 and 30$\,$K, and massive, 
350 and 570\Mo, respectively (Mangum et al. 1992; see also 
Liechti \& Walmsley 1997).
The total mass and luminosity of the region
are about $3\times 10^4$\Mo~and $5\times10^4$\Lo~(Chandler et al. 
1993a, 1993b). Star formation activity is indicated by
the presence of millimeter continuum sources (e.g., Mangum et al. 1991;
Chandler et al. 1993a), a high velocity molecular outflow 
detected in CS J=5--4 (Richardson et al. 1994)
and various species of masers (e.g., Plambeck \& Menten 1990;
Kurtz et al. 2004; Fish et al. 2005). 
Particularly interesting is the outflow traced by 
CH$_3$OH 44$\,$GHz masers: sub-arcsecond angular
resolution observations by Kogan \& Slysh (1998) 
show the presence of several CH$_3$OH 44$\,$GHz masers 
grouped in a bipolar structure. The morphology of the
blueshifted masers resembles a bow shock. Higher
sensitivity CH$_3$OH 44$\,$GHz observations by Kurtz et al.
(2004) confirm the Kogan \& Slysh (1998) results and
reveal a possible second bow shock. 

In summary, numerous observations of DR21(OH) clearly identify 
it as a very young massive star forming region, with multiple
OB-star clusters forming.
All of the clusters appear to be in a pre-ultracompact H{~\small II}
region stage. The unambiguous evidence of massive star
formation and the relatively nearby distance ($\sim 2\,$kpc; 
Odenwald \& Schwartz 1993) make DR21(OH) an ideal laboratory to 
study early stages of massive star formation.

Extensive work by several groups (e.g., see Eisl\"offel et al. 2000)
has shown that star formation regions usually contain numerous
weak centimeter continuum sources. These
sources trace a variety of astrophysical phenomena, including
thermal jets, stellar
winds, and gyro-synchrotron emission from young stellar objects.
The location and characterization of these weak radio sources has proved 
extremely valuable to our understanding of star formation phenomena
(e.g., Zapata et al. 2004; Ignace \& Churchwell 2004;
Girart et al. 2002; Reipurth et al. 2002).

Despite a wealth of molecular line and millimeter continuum observations
of DR21(OH), it has never been reliably imaged with high sensitivity
and angular resolution in the centimeter continuum. 
Relatively low quality VLA\footnote{The Very Large
Array (VLA) is operated by the National Radio Astronomy Observatory
(NRAO), a facility of the National Science Foundation operated
under cooperative agreement by Associated Universities, Inc.}
archive continuum 
observations were reported by Argon et al. (2000) and Fish et al. (2005), 
and the most sensitive cm continuum data available in the 
literature has 3$\sigma$ sensitivities of 10 -- 15 \mjyb~
(Johnston et al. 1984; Mangum et al. 1992). 
In this work we present high angular resolution 
($\theta_{syn}< 1$\arcsec) and high sensitivity ($rms < 1\,$mJy)
observations of the 
radio continuum at 3.6, 1.3, and 0.7$\,$cm, as well as observations
of the CH$_3$OH 44$\,$GHz masers in DR21(OH).

\section{Observations and Data Reduction}

We used the VLA
to conduct radio continuum observations of
DR21(OH) at X (3.6$\,$cm), K (1.3$\,$cm) and Q (7$\,$mm) bands. 
In addition, we report observations of CH$_3$OH 44$\,$GHz masers. 
Details of the observations are summarized in Table~1\footnote{The
observations reported in this paper are from the NRAO proposals AH745 and
AK586.}.

The 3.6$\,$cm and 1.3$\,$cm observations were conducted using
the standard VLA continuum mode, i.e., 50$\,$MHz bandwidth, 
4IF mode. The quasar 3C48 was used as primary (flux) calibrator
and J2007+404 was used as secondary (phase) calibrator. 
The calibration cycle of the 3.6$\,$cm and 1.3$\,$cm observations was
$\sim$ 600/60 and 220/40$\,$s 
(source/calibrator), respectively. All data reduction
was done in AIPS following standard procedures.

The 7$\,$mm continuum was
observed simultaneously with the CH$_3$OH masers using two
IF pairs. The goal of this
setup was to cross-calibrate the radio continuum observations
using the CH$_3$OH masers to minimize phase decorrelation due
to tropospheric variations. 
One pair of IFs was set to a narrow bandwidth of 3.125$\,$MHz (63
channels, 48.83$\,$kHz) to observe the CH$_3$OH 
$\nu_0 = 44.06943\,$GHz ($7_0 - 6_1, A^+$) line; 
the second pair of IFs was used with 
a bandwidth of 25.0$\,$MHz to observe the radio continuum (excluding 
the maser lines). The narrow band 
IF was centered on V$_{LSR} = 0.0$\kms. The CH$_3$OH spectra were
Hanning smoothed to a final channel width of 97.6$\,$kHz (0.66\kms).
We observed 3C286 as primary (flux) calibrator and 
1229+020 as bandpass calibration; in addition, J2007+404 was observed 
approximately every hour to correct for
phase offsets between the CH$_3$OH and continuum passbands.
Assuming the position reported in Kurtz et al.
(2004) for the brightest CH$_3$OH maser in the region, we 
self-calibrated the peak maser channel and transferred 
the calibration table to all CH$_3$OH channels, as well as
to the broadband (25$\,$MHz) continuum IFs. All calibration and
imaging was done using the NRAO package AIPS.

\section{Results \& Discussion}

We detected radio continuum emission in all three bands
(3.6$\,$cm, 1.3$\,$cm, and 7$\,$mm) as well as CH$_3$OH 
44$\,$GHz masers. Figure~1 shows the radio continuum contour maps;
Table~2 lists the parameters of the 
final continuum images and in Table~3 we report 
the observed properties of all radio continuum sources.
Figure~2 shows the combined CH$_3$OH spectrum
(measured from a 35\arcsec $\times$ 11\arcsec~box centered
between the MM1 continuum sources and 3.6$\,$cm source R1), and
Figure~3 shows the integrated intensity
(zero velocity moment) image of the CH$_3$OH emission. In Figure~4
we show the velocity field (first velocity moment) of the 
CH$_3$OH maser arcs seen in Figure~3. To facilitate comparison
between the figures, we also show in Figure~4 the 3.6$\,$cm
radio continuum emission (contours; see Figure~1).
CH$_3$OH line parameters are listed
in Table~4.

\vspace{-0.5cm}

\subsection{CH$_3$OH 44$\,$GHz Masers: Outflow in DR21(OH)}

The CH$_3$OH masers were observed in this work with a two-fold
goal: to cross-calibrate the 7$\,$mm radio continuum 
and to observe the masers with higher angular resolution 
($\theta_{syn} \sim$ 0.6\arcsec~versus 1.7\arcsec) and sensitivity 
(rms $\sim$ 5\mjyb~versus 50\mjyb) than the observations reported 
by Kurtz et al. (2004). Our observations however, have poorer 
spectral resolution than those of Kurtz et al. (2004), i.e., 
0.7 versus 0.17\kms. Given the difference in spectral and angular
resolution, as well as flux density calibration
uncertainties (see Kurtz et al. 2004), we consider that our 
data are not suitable for a variability study of the masers 
by comparing these data sets.

Figure~3 shows the integrated intensity distribution
of CH$_3$OH masers in DR21(OH). We detected more than 30 new
maser spots. Figure~4 shows the velocity distribution. 
As noted by Kurtz et al. (2004)
and Kogan \& Slysh (1998), the CH$_3$OH 44$\,$GHz masers
trace a bipolar structure, with blueshifted masers located 
predominately in the eastern lobe,
and red shifted masers located in the western lobe. 
As proposed by Kurtz et al. (2004, see also Plambeck \& Menten 1990 for
the case of 95$\,$GHz CH$_3$OH masers), this
organized maser distribution (in space and velocity) is likely tracing
an outflow in DR21(OH). Given the mass and temperature
of the MM2 core (Mangum et al. 1991), the presence of an 
east-west molecular outflow (e.g., Lai et al. 2003; Chandler et al. 1993b)
and the widespread association of 
CH$_3$OH Class I masers with young massive stars
(e.g., Ellingsen 2006; Pratap et al. 2008), 
the driving source is likely a young massive stellar object 
somewhere in MM2, located 
toward the center of the bipolar maser distribution.

The distribution of masers within each lobe is quite
intriguing. Kogan \& Slysh (1998) found that the CH$_3$OH
44$\,$GHz masers in the blueshifted group are distributed along
an arc, suggestive of a bow-shock origin. 
The higher sensitivity observations of Kurtz et al. (2004) confirm 
the morphology in the blue lobe and show the presence
of a possible second arc. Our still higher sensitivity observations
enable a more detailed characterization of the CH$_3$OH maser
distribution in DR21(OH).

Our data confirm the 
two-arc maser distribution of the eastern (blue) lobe 
and reveal that 
the masers in the {\it western} lobe are also
grouped in two arc-like structures. As explicitly suggested by 
thin lines in Figure~4, the CH$_3$OH masers appear to trace 
two nested bow-shocks in the outflow. This distribution
is remarkably similar to the jet bow-shock outflow model shown by
Arce et al. (2007). The occurrence of sequential 
bow-shocks (outflow events) is also reminiscent
of the morphology of the outflow in DR21 (Smith et al. 2006).  
Clearly, the sky distribution of the CH$_3$OH masers traces 
a 2D projection of a 3D structure; the masers might originate in 
a cut through parabolic-like outflow lobes where the velocity 
coherence and beaming are favorable for maser amplification toward
our line of sight.\footnote{We thank P. T. P. Ho for stressing 
this point.}

Given the symmetric bow-shock morphology and the low velocity 
gradient between the red and blueshifted 
maser groups ($<10$\kms, Figure~4), 
it appears that the outflow is almost in the plane of the sky.
Assuming that the CH$_3$OH masers are tracing two individual (nested) 
outflow events, and assuming, for simplicity, a constant expansion 
velocity of 100\kms~for each
lobe (e.g., as observed in H$_2$O masers in IRAS$\,$20126+4104,
Moscadelli et al. 2005), then the dynamical ages 
are $\sim 900\,$yr and $\sim 1500\,$yr for the inner and outer shocks, 
respectively (a distance of 2$\,$kpc was assumed), i.e., the lower
limit of the outflow age is $\sim 1500\,$yr. Thus, the outflow 
could be quite young --- younger than the `pulsed' outflow in 
Cep-A (e.g., Bally 2008). However, we note that high angular resolution
(VLBI) proper motion studies are required to determine the
velocity of the CH$_3$OH masers in the region.
In particular, CH$_3$OH molecules may not survive in high 
velocity J-shock environments (e.g., Garay et al. 2002).
Alternatively, the CH$_3$OH masers may be tracing a slower
shock traveling along the high speed flow and/or the working surface 
between two shocks.\footnote{We acknowledge E. Churchwell and 
J. Cant\'o for pointing out these possibilities.}

A complication to the single-outflow
interpretation of the CH$_3$OH masers in DR21(OH) is the 
misalignment between the arcs, particularly with respect to 
the eastern arc (Figure~4; i.e.,
a change in position angle of $\sim$15\arcdeg~east of north).
Such a change in position angle could result from jet precession
combined with inhomogeneities in the molecular gas that is
interacting with the flow, similar to the large scale 
outflow in G192.16$-$3.82 (Devine et al. 1999). Nevertheless, 
despite the misalignment, the outflow seems to be quite
collimated, with an apparently constant opening cross-section at 
scales greater than 10$^4\,$AU ($\sim$0.05$\,$pc) from the 
symmetry center.

\subsection{Radio Continuum: A Cluster of Radio Sources}

In order to investigate the driving source of the CH$_3$OH maser
outflow in DR21(OH), we conducted sub-arcsecond 
radio continuum observations at 3.6$\,$cm, 1.3$\,$cm, and
7$\,$mm (Figure~1). We detected eight radio continuum
sources (see Table~3). In Figure~4 we show the CH$_3$OH
velocity field superimposed on the 3.6$\,$cm continuum 
map. One of the 3.6$\,$cm continuum sources (R1) is located close
to the projected center of the CH$_3$OH maser outflow; it may mark the
position of the massive (proto)star that drives the outflow.

Alternatively, DR21(OH)-R1 could mark the position of
a young low-mass star. The radio emission could be
gyro-synchrotron radiation from an active magnetosphere, 
just as the low mass companion of 
$\theta^1$ Ori A (Felli et al. 1993) or the variable radio
source in IRAS$\,$20126+4104 (Hofner et al. 2007). This 
hypothesis is supported by the detection of other 
weak radio sources (R5, R6 and possibly R2; Figure~1), that
together with R1 may represent the high end of the (radio) luminosity 
function of a cluster of low mass objects forming 
among the young massive stars.

In Table~5 we list
the spectral index between 3.6 and 1.3$\,$cm of all sources
detected at both wavelengths. The spectral indexes of
R1, R2, R5, and R6 (a different interpretation of the radio
emission from R3 and R4 is given below) are 
between $-0.6$ and 0.4, which is similar to 
the range of spectral index variability of
the low mass star in $\theta^1$ Ori A (Felli et al. 1993).
Further observations of the weak population of radio sources
in DR21(OH) are necessary to assess whether the radio emission
is variable as expected from the gyro-synchrotron 
hypothesis, and with this, to study how 
low-mass pre-main-sequence stars (class I/II) and very 
young class 0 high-mass objects 
can coexist considering their clearly different 
evolutionary timescales.

In contrast to the radio emission from 
R1, R5, and R6 (and possibly R2), the 
radio sources MM1-NW, MM1-SE, R3 and 
R4 appear to be related to the massive
young stellar object in MM1. We propose that 
these radio sources are tracing 
shock-ionized gas in the jet
from a young massive star. We prefer
this interpretation because: 

\noindent 1. MM1-NW, SE, R3 and R4 are collinear 
(southeast-northwest orientation), which 
suggests a common driving source.

\noindent 2. In principle the similar flux densities
of MM1-NW and SE could be explained by 
radio emission from two hypercompact 
H{~\small II} regions that are in a very similar
evolutionary state. 
However, the morphology of the two radio sources is also similar;
after deconvolution, both sources appear to be elongated at
approximately the same position angle ($\sim 40$\arcdeg; see
Table 3).  This position angle is almost perpendicular to the
MM1-NW to R4 orientation (i.e., orthogonal to the putative jet;
see Figure~1), and may arise from a bow shock aligned with the
MM1-NW to R4 axis.  The elongation would result from our edge-on
view of the shocks, corresponding to the broad tail of a
relatively
large-opening-angle bow shock, emitting free-free radiation.
Given the symmetry, we consider it more likely that MM1-NW
and SE are manifestations of a single radio jet.

\noindent 3. The spectral indexes between X and K bands
are very similar for MM1-NW and SE ($\alpha \sim 0.8$); 
in addition, the spectral index of R4 is also similar to
those of MM1-NW and SE within the errors (Table~5).
A spectral index of $\sim 0.8$ is consistent with
thermal emission from finite stratified density
gas being ionized by UV photons from the cooling 
region of a shock (e.g., Ghavamian \& Hartigan 1998).

The weak radio emission supports the interpretation that the most
massive stellar object in MM1 is a high mass protostellar object 
(HMPO; e.g., Beuther et al. 2007).

In Figure~1 we show the location of 6.7$\,$GHz CH$_3$OH
masers (Harvey-Smith et al. 2008) and H$_2$O masers
(Palmer \& Goss, {\it in prep.}). The H$_2$O masers are distributed
throughout the region, and thus are unlikely to be driven
by a single young stellar object. Two of the weak 
radio sources (R1 and R6) may be associated with
H$_2$O masers, and another H$_2$O maser is found
almost at the center of MM1-NW/SE. The detection
of a H$_2$O maser at the center of the double radio
source is reminiscent of IRAS$\,$20126+4104 (Hofner et al. 2007;
Moscadelli et al. 2005)
where a clump of H$_2$O masers is also located toward
the center of a double radio source. The radio continuum emission
associated with the H$_2$O masers in IRAS$\,$20126+4104
is likely due to shock-ionized gas in the jet from a young massive star
(see also $\S 3.5$).

\subsection{The Nature of the 7$\,$mm Emission}

As evident from Figure~1, the 7$\,$mm source has the
elongation expected from superposition of emission
from MM1-NW, SE, R2, R3, and R4 when observed with lower
angular resolution. The southeast -- northwest
elongation is also consistent with the 2$\,$cm continuum
map shown by Fish et al. (2005).
To further investigate the nature of the 
7$\,$mm source we convolved the 3.6 and 1.3$\,$cm 
observations to the synthesized beam of the 7$\,$mm map. 
In Figure~5 we show the (radio) flux density distribution 
of MM1 after convolution, i.e., 
MM1-NW + SE + R2 + R3 + R4. 
The 7$\,$mm flux density point lies slightly
above the power law fit.  This indicates
that at 7mm the flux is dominated by free-free emission with a small
($\sim 1.4\,$mJy or 10$\%$ of the total flux density)
excess, most likely due to thermal dust emission.

\subsection{Infrared Counterparts}

The DR21/W75 region (including DR21(OH)) has been extensively
studied in the infrared (e.g., Davis et al. 2007; Kumar et al. 2007;
Smith et al. 2006; Marston et al. 2004). In Figure~6 we show
the {\it Spitzer} IRAC image of the region 
(3.6$\,\mu$m blue; 4.5$\,\mu$m green; 
8.0$\,\mu$m red)\footnote{Data were retrieved from the 
{\it Spitzer} archive (http://irsa.ipac.caltech.edu/applications/Spitzer/Spitzer/), 
project PID-1021, ``DR21 and its 
Molecular Outflow'', Marston et al. (2004).}. These
IRAC observations have been discussed in a number of papers 
(e.g., Kumar et al. 2007) and are presented here
to facilitate comparison between the CH$_3$OH, radio continuum 
and infrared data.

Figure~6 shows a number of sources with 4.5$\,\mu$m
excess particularly along the large scale north-south
(dark) filament (see Kumar et al. [2007] for a larger scale 
view of the region). Excess in the 4.5$\,\mu$m band is a known 
tracer of shocks (e.g., Smith et al. 2006; 
Cyganowski et al. 2008), thus, Figure~6 exemplifies that star
formation in DR21(OH) is not limited to the mm cores but is 
found throughout the region. The eastern CH$_3$OH arcs may
be associated with 4.5$\,\mu$m excess emission, however there 
is no clear evidence for 4.5$\,\mu$m excess toward the western arcs.

Shocked gas in the region has also been studied by Davis et al. (2007)
based on H$_2$ 2.12$\,\mu$m observations. These authors 
did not find H$_2$ emission corresponding to the CH$_3$OH masers.
This might be due to the high optical depth at 2.12$\,\mu$m for these 
deeply embedded regions in the vicinity of MM1 and MM2. 
However, slightly outside the area shown in Figure~4, Davis et al. (2007) 
found two H$_2$ regions that are somewhat 
aligned with the maser outflow (B7-1 and B5-1; see 
Table 3 and Fig. A.2 of Davis et al. [2007]).

The MM1 radio continuum sources are almost coincident with the peak of
an extended ``red'' object (see Figure~6); the infrared 
counterpart of MM1 was also detected at 24$\,\mu$m (saturated) and 
70$\,\mu$m with MIPS. Figure~6 also shows a ``blue'' star
located very close to the symmetry center of the 
CH$_3$OH outflow. The star is clearly detected at
3.6 and 4.5$\,\mu$m (brighter at 3.6$\,\mu$m) and tentatively
detected at 5.8$\,\mu$m blended with the MM1 emission.
The star is also detected in J, H, and K bands
(2MASS), and in optical DSS images. 

The near-infrared (NIR) colors of the
``blue'' object are intriguing. The 2MASS colors are 
$J-H$ = 0.32 and $H-K_s$ = 0.11, 
which, if placed into a NIR color-color
diagram (e.g., Hillenbrand et al. 1998, 
Meyer et al. 1997, Marston et al. 2004, 
Apai et al. 2005), are consistent with an 
early F-type main sequence star with $\sim$1 mag of 
optical extinction\footnote{More
specifically, one 
magnitude of optical extinction corresponds to extinctions of 
0.28 mag in J band, 0.18 mag in H band, and 0.11 mag in K 
band (Mathis 1990). Correcting the 2MASS colors for these extinction
values and based on Ducati et al (2001), the colors are consistent
with those of an F0 - F2 star.}. If it is a F0 star, 
then to attain V = 13 mag (as roughly observed), such a star 
would be at a distance of 
$\sim$700$\,$pc, and hence would not be associated with 
the DR21 region. However, the {\it Spitzer} colors are inconsistent with 
this interpretation. If the optical extinction were 1 mag, a 
{\it Spitzer} IRAC color of [3.6]-[4.5] = 0.53
could not be caused by extinction, 
because this color indicates much higher extinction values ($A_V > $10 mag).
Hence, the blue object could still be associated with 
the DR21 region, and other circumstances (unresolved multiplicity, existence 
of a transitional circumstellar disk, etc.) might govern the infrared colors. 
Based on the current data, it is hard to decide whether the blue object 
is the driving source for the CH$_3$OH outflow. 
An in-depth investigation of the nature
of this near-IR source is beyond the scope of this work, and 
should be the topic of a future study.

\subsection{A New Observational Phase during the 
Formation of Massive Stars}

It is interesting to compare the characteristics 
of the radio emission in DR21(OH) with 
that of other young massive stellar objects, in particular 
G31.41$+0.31$ and IRAS$\,$20126+4104 (Araya et al. 2003, 
2008; Hofner et al. 2007). 
In all three cases, there are multiple radio continuum sources
toward the center of molecular cores. Some of the continuum
sources are consistent with 
radio emission from shock-ionized gas 
and not due to direct photo-ionization from an embedded massive
star (see for example figure~1 of Hofner et al. 2007). 
The intermediate mass object GGD27 (HH80/81) is a clear example
of ionized knots moving along a collimated jet as shown by proper 
motion studies (Mart\'{\i} et al. 1995, 1998).

We propose that there is a period during the formation of
massive stars when (possibly recurrent) jet events interact
with the molecular core and ionize the medium via shocks.
Such a process could produce the double radio sources
we have detected in the core of several young
massive star forming regions. The ionization process would 
eventually become dominated by the development of hypercompact and 
then ultracompact H{~\small II} regions.

\subsection{Further Implications}

The data reported in this paper show regularities in
the orientation of outflows among themselves and also 
with respect to the large scale distribution of dust and gas 
in the region. 
The angular distance between the MM1 radio emission and the 
geometrical center of the CH$_3$OH outflow (see Figure~4)
is too large to assume that the driving source
of the outflow is located in MM1. Thus, our data 
indicate the presence of two distinct young massive
stellar objects (or systems) with nearly parallel outflows/jets;
i.e., both the putative MM1-NW to R4 jet and the CH$_3$OH outflow have 
similar position angles ($\sim 130^\circ$).
This suggests that the orientation of jets/outflows is, to
first order, determined by the large scale properties of the parent
cloud. This conclusion is further supported by 
the powerful outflow from DR21 which also has
a dominant east-west orientation (e.g., Marston
et al. 2004). The statistically significant orientation of the
outflows in an east-west direction was also
discussed by Davis et al. (2007). Other examples of quasi-parallel 
outflow from different young massive stellar objects
within the same star forming region include
Cep-A and IRAS$\,$05358+3543 (Beuther et al. 2002; Bally 
2008). 

The complete DR21 to DR21(OH) region is part of 
a long molecular cloud filament (or sheet seen edge-on) 
oriented north-south (e.g., Marston et al. 2004; Motte et al. 2005, 2007). 
The outflows are mostly perpendicular to the large
scale north-south filament, and approximately aligned with 
the large scale direction
of the magnetic field through the filament, which, 
as discussed by Vall\'ee \& Fiege (2006), shows a predominant
east-west orientation (see also Lai et al. 2003). 
If the outflows arise from (proto)stars with accretion disks,
then the disks would tend to be oriented edge-on along
the filament, with disk symmetry axes oriented 
east-west.\footnote{A potential counterexample is
ERO3 in the DR21(OH)N region, where Harvey-Smith et al. (2008)
and Harvey-Smith \& Soria-Ruiz (2008) report a possible Keplerian
disk traced by CH$_3$OH masers and oriented east-west. However, we
consider that future studies may be needed to confirm their result.}

\section{Summary} 

We report sub-arcsecond observations of the radio
continuum at 3.6, 1.3, and 0.7$\,$cm, and also
 of the CH$_3$OH 44$\,$GHz masers in the 
DR21(OH) region. We detected a cluster of radio
sources; the two brightest sources are located
toward the molecular core MM1 and appear to trace
radio emission from shock-ionized gas.

We detected more than 30 new CH$_3$OH 44$\,$GHz
maser spots in DR21(OH). Our observations confirm and
delineate a bow-shock distribution
of the masers. The masers show a well-separated (in space
and velocity) bipolar outflow almost symmetrically located
with respect to the MM2 core, which is likely to harbor the
driving source of the flow. The CH$_3$OH masers appear to trace 
two different outburst events. 
The sequential bow-shock morphology 
observed in DR21(OH) is reminiscent of that of the 
DR21 outflow (Smith et al. 2006).

The driving source of the CH$_3$OH outflow in DR21(OH) is unlikely
to be the one responsible for the radio emission in MM1.
More likely, two different young massive stellar
objects are driving outflows/jets at very similar
position angles. At even greater scales, the prominent 
outflow from DR21 also has an $\sim$east-west orientation. 
As also noted by Davis et al. (2007),
throughout the large scale filament in the DR21-DR21(OH) region
we find that the outflows/jets are quite parallel in
a $\sim$ east-west direction, which is 
perpendicular to the (north-south) filament. This suggests
that the accretion disks form mostly edge-on along the large scale 
filament, i.e., disk symmetry axis orthogonal to the filament elongation.

\acknowledgments

E.A. acknowledges support from an University of New Mexico 
postdoctoral fellowship and a National Radio Astronomy Observatory
(NRAO) Jansky fellowship. We thank L. Deharveng for a
discussion regarding the IR emission in DR21(OH) and
an anonymous referee for a thorough review and comments that 
improved the manuscript. We also 
thank P. Palmer and M. Goss for providing the locations of the H$_2$O
masers. This research made use of the NASA's Astrophysics Data System, 
and archival data from the {\it Spitzer Space Telescope} and 
the Digitized Sky Survey. The Digitized Sky Surveys were 
produced at the Space Telescope Science Institute under U.S. 
Government grant NAG W-2166.
The Two Micron All Sky Survey catalog was also consulted as part 
of this investigation; 2MASS is a joint project of the University 
of Massachusetts and the Infrared Processing and Analysis 
Center/California Institute of Technology, funded by NASA and 
the National Science Foundation.

\clearpage

\begin{deluxetable}{lcccc}
\tabletypesize{\scriptsize}
\tablecaption{VLA Observations \label{tbl-1}}
\tablewidth{0pt}
\tablehead{
\colhead{Parameter}& \colhead{X-Band (3.6$\,$cm)}  & \colhead{K-Band (1.3$\,$cm)}& \colhead{Q-Band (7$\,$mm)$^a$} & \colhead{CH$_3$OH 44$\,$GHz$^a$} }
\startdata
{\rm Date}                & Nov 16, 2004           & May 05, 2005                & Sep 07, 2001                 & Sep 07, 2001           \\
VLA Configuration         & VLA-A                  & VLA-B                       & VLA-C                        & VLA-C                  \\
RA$^b$	                  & 20 39 01.000           & 20 39 01.000                & 20 39 00.900                 & 20 39 00.900           \\   
Dec$^b$                   & 42 22 40.00            & 42 22 40.00                 & 42 22 47.00                  & 42 22 47.00            \\
$\nu_o\,$(GHz)            & 8.46                   & 22.4                        & 44.1                         & 44.06943               \\
Correlator Mode           & 4IF Cont.              & 4IF Cont.                   & 4IF Line                     & 4IF Line               \\
Bandwidth per IF (MHz)    & 50$\,$MHz              & 50$\,$MHz                   & 25.0 (2 Pol)                 & 3.125 (2 Pol)          \\
Flux Density Calib.       & 3C48                   & 3C48                        & 3C286                        & 3C286                  \\
~~~~~Assumed S$_\nu$ (Jy) & 3.15                   & 1.12                        & 1.43                         & 1.43                   \\           
Phase Calib.              & J2007+404               & J2007+404                    & J2007+404$^c$                 & J2007+404$^c$           \\
~~~~~Measured S$_\nu$ (Jy)& 2.54$\pm$0.01          & 1.78$\pm$0.02               & 1.34$\pm$0.04                & 1.35$\pm$0.04          \\
\enddata
\tablenotetext{a}{The Q-Band (7$\,$cm) and CH$_3$OH 44$\,$GHz observations were conducted simultaneously using a
4IF mode. Self-calibration solutions from the CH$_3$OH maser observations 
were used to calibrate the Q-Band continuum.}
\tablenotetext{b}{Phase tracking center (J2000).}
\tablenotetext{c}{Used to correct phase offsets between the wideband and 
CH$_3$OH spectral line band.}
\end{deluxetable}

\clearpage

\begin{deluxetable}{lccc}
\tabletypesize{\scriptsize}
\tablecaption{Parameters of the Radio Continuum Images \label{tbl-2}}
\tablewidth{0pt}
\tablehead{
\colhead{Parameter}       &  \colhead{X-Band (3.6$\,$cm)}  & \colhead{K-Band (1.3$\,$cm)}   & \colhead{Q-Band (7$\,$mm)} }
\startdata
Syn. Beam                 & 0\ads20 $\times$ 0\ads19       & 0\ads25 $\times$ 0\ads25       & 0\ads92 $\times$ 0\ads74  \\
Syn. Beam P.A.            & $-$18.2\ad                     & 62.1\ad                        & $-$74.9\ad                \\
rms ($\mu$Jy$\,$b$^{-1}$) & 19                             & 36                             & 400                       \\
\enddata
\end{deluxetable}

\clearpage

\begin{deluxetable}{lcccccc}
\tabletypesize{\scriptsize}
\tablecaption{Radio Continuum Sources$^a$ \label{tbl-2}}
\tablewidth{0pt}
\tablehead{
\colhead{Source} & \colhead{Band} & \colhead{RA(J2000)} & \colhead{Dec(J2000)} & \colhead{Size$^b$, P.A.} & \colhead{I$_{\nu}$} & \colhead{S$_{\nu}$} \\
\colhead{      } & \colhead{    } & \colhead{20\h 39\m} & \colhead{42\ad 22\am}& \colhead{(AU,\arcdeg)} & \colhead{(\mjyb)} & \colhead{(mJy)}}  
\startdata
DR21(OH)-R1      & 3.6$\,$cm  & 00\sec612(1)  & 43\ads44(1)  & 440$\times$100,   86\ad &   0.17(2)   &  0.26(4) \\  
                 & 1.3$\,$cm  & 00\sec606(4)  & 43\ads45(3)  & 670$\times <$160, 59\ad &   0.14(4)   &  0.20(8) \\
MM1-NW           & 3.6$\,$cm  & 00\sec9763(1) & 48\ads980(2) & 290$\times <$50,  34\ad &   1.14(2)   &  1.40(4) \\
                 & 1.3$\,$cm  & 00\sec9765(1) & 48\ads981(1) & 200$\times$80,    37\ad &   2.70(4)   &  2.95(7) \\
MM1-SE           & 3.6$\,$cm  & 01\sec0079(1) & 48\ads707(2) & 280$\times$155,   46\ad &   1.18(2)   &  1.55(4) \\
                 & 1.3$\,$cm  & 01\sec0079(1) & 48\ads695(1) & 235$\times$225,   51\ad &   2.83(4)   &  3.43(7) \\
MM1$^c$          &   7$\,$mm  & 01\sec02$^d$  & 48\ads8$^d$  &1830$\times <$460, 123\ad&   9.4(4)    &  13.4(8) \\
DR21(OH)-R2      & 3.6$\,$cm  & 01\sec042(1)  & 48\ads93(1)  & 330$\times <$100, 65\ad &   0.16(2)   &  0.18(4) \\   
                 & 1.3$\,$cm  & 01\sec045(3)  & 48\ads91(2)  & 440$\times <$300, 70\ad &   0.16(4)   &  0.20(7) \\
DR21(OH)-R3      & 1.3$\,$cm  & 01\sec054(2)  & 48\ads43(2)  & 690$\times <$300, 111\ad&   0.19(3)   &  0.32(9) \\
DR21(OH)-R4      & 3.6$\,$cm  & 01\sec083(2)  & 48\ads35(2)  & 465$\times$420,   94\ad &   0.14(2)   &  0.31(6) \\   
                 & 1.3$\,$cm  & 01\sec077(3)  & 48\ads32(2)  & 745$\times$315,   97\ad &   0.21(3)   &  0.45(10)\\  
DR21(OH)-R5      & 3.6$\,$cm  & 01\sec303(1)  & 49\ads019(8) & 285$\times <$135, 77\ad &   0.18(2)   &  0.21(4) \\   
                 & 1.3$\,$cm  & 01\sec308(2)  & 49\ads05(2)  & $<230\times <90$, 144\ad&   0.16(4)   &  0.12(5) \\
DR21(OH)-R6      & 3.6$\,$cm  & 01\sec337(1)  & 52\ads00(2)  & 190$\times <$310, 60\ad &   0.11(2)   &  0.12(4) \\   
                 & 1.3$\,$cm  & 01\sec343(2)  & 51\ads98(2)  & 180$\times <100$, 125\ad&   0.20(4)   &  0.18(6) \\
\enddata
\tablenotetext{a}{ Parameters obtained from a 2D Gaussian fit of the brightness distribution using the task {\tt JMFIT} in AIPS.
The format of the error notation is, e.g., xx.xxx(y) = xx.xxx $\pm$ 0.00y.}
\tablenotetext{b}{ Nominal deconvolved size (major and minor axes, and position angle). A distance of 2$\,$kpc was assumed
(Odenwald \& Schwartz 1993). We list a quarter of the synthesized beam (0\ads05$\approx$100$\,$AU at X-Band) 
as upper limit of the size in the case of unresolved sources
for which JMFIT does not report a limit.}
\tablenotetext{c}{ The sources MM1-NW, MM1-SE, DR21(OH)-R2, R3, and R4 are blended at 7$\,$mm.}
\tablenotetext{d}{ For the self-calibration, we assumed the position of the brightest CH$_3$OH maser as given by Kurtz et al. (2004). Thus, the
position error of our Q-band observations is dominated by the Kurtz et al. (2004) astrometric
accuracy of $\sim$0\ads6.}
\end{deluxetable}

\clearpage

\begin{deluxetable}{lccccccc}
\tabletypesize{\scriptsize}
\tablecaption{CH$_3$OH Line Parameters \label{tbl-2}}
\tablewidth{0pt}
\tablehead{
\colhead{Maser ID} & \colhead{RA (J2000)} & \colhead{Dec (J2000)} & \colhead{I$_\nu^a$} & \colhead{V$_{LSR}$} & \colhead{$\Delta$V$_{3\sigma}^a$} 
& \colhead{$\int I_{\nu} dv$} & \colhead {Notes$^b$} \\
\colhead{} & \colhead{} & \colhead{} & \colhead{(Jy$\,$b$^{-1}$)} & \colhead{(\kms)} & \colhead{(\kms)} & \colhead{(Jy$\,$b$^{-1}$ \kms)} }
\startdata
CH$_3$OH-1 & 20 38 58.68 & 42 22 22.1 & 0.40   &  $-$1.7   &  2.0 & 0.41 & N\\  
CH$_3$OH-2 & 20 38 58.92 & 42 22 27.2 & 0.39   &  $-$6.3   &  3.3 & 0.92 & N\\  
CH$_3$OH-3 & 20 38 59.29 & 42 22 48.8 &76.8    &     0.3   &  3.3 & 79.3 & Y-1\\
CH$_3$OH-4 & 20 38 59.32 & 42 22 49.6 & 2.95   &     1.0   &  2.0 & 3.28 & N\\  
CH$_3$OH-5 & 20 38 59.35 & 42 22 47.4 & 2.05   &  $-$1.7   &  2.6 & 2.68 & Y-2\\
CH$_3$OH-6 & 20 38 59.56 & 42 23 05.3 & 0.49   &  $-$1.0   &  2.0 & 0.57 & Y-3\\
CH$_3$OH-7 & 20 38 59.64 & 42 22 47.2 & 0.092  &  $-$0.3   &  2.6 & 0.16 & N\\  
CH$_3$OH-8 & 20 38 59.72 & 42 22 49.6 & 0.27   &  $-$0.3   &  2.6 & 0.29 & N\\  
CH$_3$OH-9 & 20 38 59.73 & 42 23 16.2 & 0.27   &  $-$0.3   &  2.6 & 0.49 & N\\  
CH$_3$OH-10& 20 38 59.73 & 42 22 47.6 & 0.12   &  $-$0.3   &  1.3 & 0.12 & N\\  
CH$_3$OH-11& 20 38 59.76 & 42 22 45.5 & 0.23   &  $-$0.3   &  2.6 & 0.37 & N\\  
CH$_3$OH-12& 20 38 59.83 & 42 22 50.1 & 0.14   &  $-$1.7   &  3.3 & 0.23 & N, O\\
CH$_3$OH-13& 20 38 59.89 & 42 22 46.0 & 3.01   &  $-$0.3   &  3.3 & 3.93 & Y-4, O\\
CH$_3$OH-14& 20 38 59.92 & 42 22 47.9 & 0.12   &  $-$0.3   &  3.3 & 0.18 & N\\  
CH$_3$OH-15& 20 38 59.92 & 42 22 50.3 & 0.21   &  $-$1.0   &  3.3 & 0.28 & N, O\\
CH$_3$OH-16& 20 38 59.93 & 42 22 45.0 & 5.27   &     0.3   &  2.6 & 6.07 & Y-6\\ 
CH$_3$OH-17& 20 38 59.99 & 42 22 35.2 & 0.16   &  $-$2.3   &  1.3 & 0.14 & N\\  
CH$_3$OH-18& 20 39 00.08 & 42 22 47.1 & 0.26   &  $-$1.7   &  2.0 & 0.38 & N\\  
CH$_3$OH-19& 20 39 00.09 & 42 22 44.7 & 0.14   &  $-$8.3   &  1.3 & 0.16 & N\\  
CH$_3$OH-20& 20 39 00.12 & 42 22 43.7 & 0.38   &  $-$5.6   &  2.0 & 0.42 & N\\  
CH$_3$OH-21& 20 39 00.13 & 42 22 47.1 & 0.43   &  $-$0.3   &  2.6 & 0.83 & N\\  
CH$_3$OH-22& 20 39 00.14 & 42 22 48.8 & 0.081  &  $-$3.0   &  3.3 & 0.19 & N\\  
CH$_3$OH-23& 20 39 00.19 & 42 22 47.7 & 3.26   &  $-$1.7   &  2.0 & 3.48 & Y-7\\
CH$_3$OH-24& 20 39 00.26 & 42 22 46.0 &21.3    &  $-$0.3   &  3.3 & 25.7 & Y-8\\
CH$_3$OH-25& 20 39 00.29 & 42 22 47.4 & 5.74   &  $-$0.3   &  3.9 & 7.10 & N\\  
CH$_3$OH-26& 20 39 00.33 & 42 22 48.5 & 0.15   &  $-$1.0   &  2.0 & 0.23 & N, O\\
CH$_3$OH-27& 20 39 00.49 & 42 22 46.3 & 0.097  &  $-$3.0   &  2.0 & 0.14 & N\\  
CH$_3$OH-28& 20 39 00.55 & 42 22 47.7 & 1.08   &     1.7   &  2.6 & 1.47 & Y-9\\
CH$_3$OH-29& 20 39 00.60 & 42 22 45.0 & 0.12   &  $-$2.3   &  2.0 & 0.20 & N\\  
CH$_3$OH-30& 20 39 01.05 & 42 22 41.6 & 0.29   &  $-$1.7   &  3.3 & 0.40 & N\\  
CH$_3$OH-31& 20 39 01.05 & 42 22 17.6 & 0.34   &  $-$3.7   &  2.0 & 0.37 & N\\  
CH$_3$OH-32& 20 39 01.09 & 42 22 01.0 & 2.21   &  $-$2.3   &  3.3 & 2.51 & N\\  
CH$_3$OH-33& 20 39 01.17 & 42 22 06.8 & 0.59   &  $-$1.0   &  2.6 & 0.82 & N\\  
CH$_3$OH-34& 20 39 01.18 & 42 22 06.8 & 0.88   &  $-$3.7   &  3.3 & 1.41 & N\\  
CH$_3$OH-35& 20 39 01.22 & 42 22 41.2 & 0.58   &  $-$3.0   &  2.6 & 0.75 & Y-10\\ 
CH$_3$OH-36& 20 39 01.47 & 42 23 06.3 & 0.064  &  $-$2.3   &  1.3 & 0.073 & N, O\\
CH$_3$OH-37& 20 39 01.48 & 42 22 07.6 & 0.78   &  $-$4.3   &  2.0 & 0.96 & N\\  
CH$_3$OH-38& 20 39 01.50 & 42 22 45.2 & 0.66   &  $-$3.0   &  2.0 & 0.85 & Y-11\\
CH$_3$OH-39& 20 39 01.51 & 42 22 40.8 & 4.68   &  $-$5.0   &  3.3 & 6.07 & Y-12\\
CH$_3$OH-40& 20 39 01.54 & 42 22 02.8 & 0.76   &  $-$3.7   &  2.0 & 1.00 & N\\  
CH$_3$OH-41& 20 39 01.67 & 42 22 43.4 & 2.50   &  $-$6.3   &  2.6 & 3.05 & Y-13\\
CH$_3$OH-42& 20 39 01.70 & 42 22 45.2 & 0.30   &  $-$5.0   &  1.3 & 0.38 & N\\  
CH$_3$OH-43& 20 39 01.76 & 42 22 45.3 & 0.21   &  $-$3.7   &  1.3 & 0.20 & N, T\\
CH$_3$OH-44& 20 39 02.02 & 42 22 41.2 & 2.23   &  $-$4.3   &  1.3 & 2.62 & Y-14, O\\
CH$_3$OH-45& 20 39 02.02 & 42 22 40.8 & 1.94   &  $-$5.6   &  1.3 & 1.82 & N, O\\  
CH$_3$OH-46& 20 39 02.08 & 42 22 45.2 & 6.29   &  $-$3.7   &  2.6 & 6.45 & Y-15\\  
CH$_3$OH-47& 20 39 02.21 & 42 22 43.4 & 0.32   &  $-$4.3   &  2.0 & 0.30 & N, O\\  
CH$_3$OH-48& 20 39 02.22 & 42 22 42.6 & 1.82   &  $-$3.7   &  2.0 & 1.80 & N, O\\  
CH$_3$OH-49& 20 39 02.24 & 42 22 44.0 & 8.77  &  $-$5.0   &  2.6 & 12.5 & Y-17, O\\
\enddata
\tablenotetext{a}{ The typical rms in one channel is 5\mjyb~for those channels with no strong ($>1\,$Jy) masers.
The values reported in this table have been corrected for
primary beam attenuation. $\Delta V_{3\sigma}$ is the
full linewidth at 3$\sigma$ level.}
\tablenotetext{b}{ We mark with `Y' or `N' whether the maser was previously reported by Kurtz et al. (2004). 
The number after `Y' is the maser number as reported by Kurtz et al. (2004).
We denote with `O' maser features that are blended in frequency with other maser emission.
Tentative detections are marked with `T'.}
\end{deluxetable}

\clearpage

\begin{deluxetable}{lc}
\tabletypesize{\scriptsize}
\tablecaption{Spectral Index between X and K Bands}
\tablewidth{0pt}
\tablehead{\colhead{Source} & \colhead{$\alpha$} }
\startdata
MM1-NW           & 0.8$\pm$0.2      \\
MM1-SE           & 0.8$\pm$0.2      \\
DR21(OH)-R1      & $-$0.3$\pm$0.2   \\ 
DR21(OH)-R2      & 0.1$\pm$0.2      \\
DR21(OH)-R4      & 0.4$\pm$0.3      \\
DR21(OH)-R5      & $-$0.6$\pm$0.2   \\
DR21(OH)-R6      & 0.4$\pm$0.2      \\
\enddata
\tablenotetext{~}{ Notes: Spectral index~($S_{\nu}\varpropto\nu^\alpha$) between 3.6 and 1.3$\,$cm 
using the values reported in Table~3.}
\end{deluxetable}

\clearpage

\begin{figure}
\includegraphics{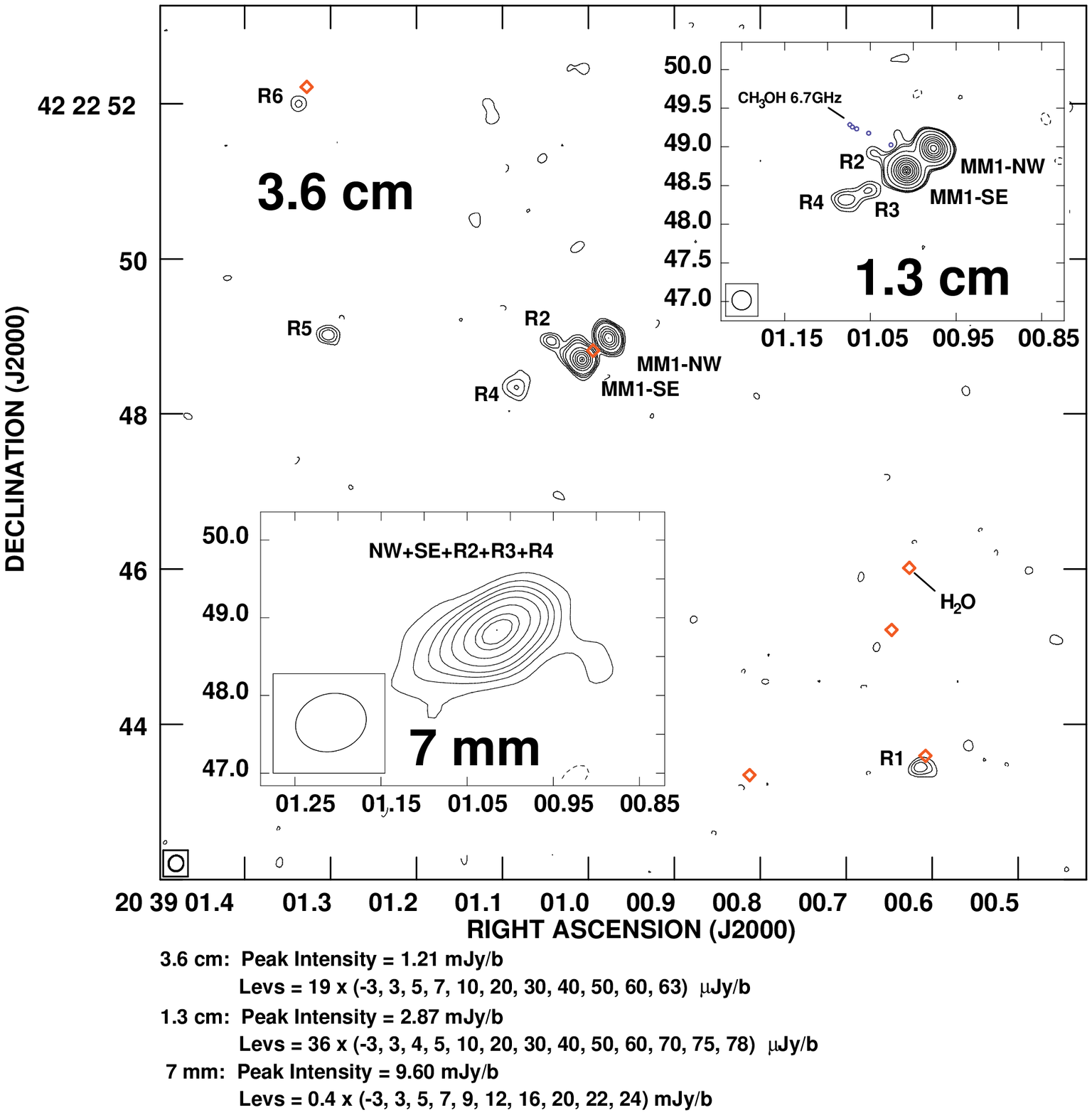} 
\vspace*{18.1cm}\caption{Radio continuum emission in DR21(OH)
at 3.6$\,$cm (X-band), 1.3$\,$cm (K-band) and 7$\,$mm (Q-band).
The synthesized beams are shown in the lower left corner of the
maps (see Table~2). The location of the 6.7$\,$GHz CH$_3$OH
masers from Harvey-Smith et al. (2008) are shown with circles in
the 1.3$\,$cm continuum map; H$_2$O masers 
are shown with diamonds in the 3.6 cm continuum map
(Palmer \& Goss, {\it in prep.}).}
\label{f1}
\end{figure}

\clearpage

\begin{figure}
\includegraphics{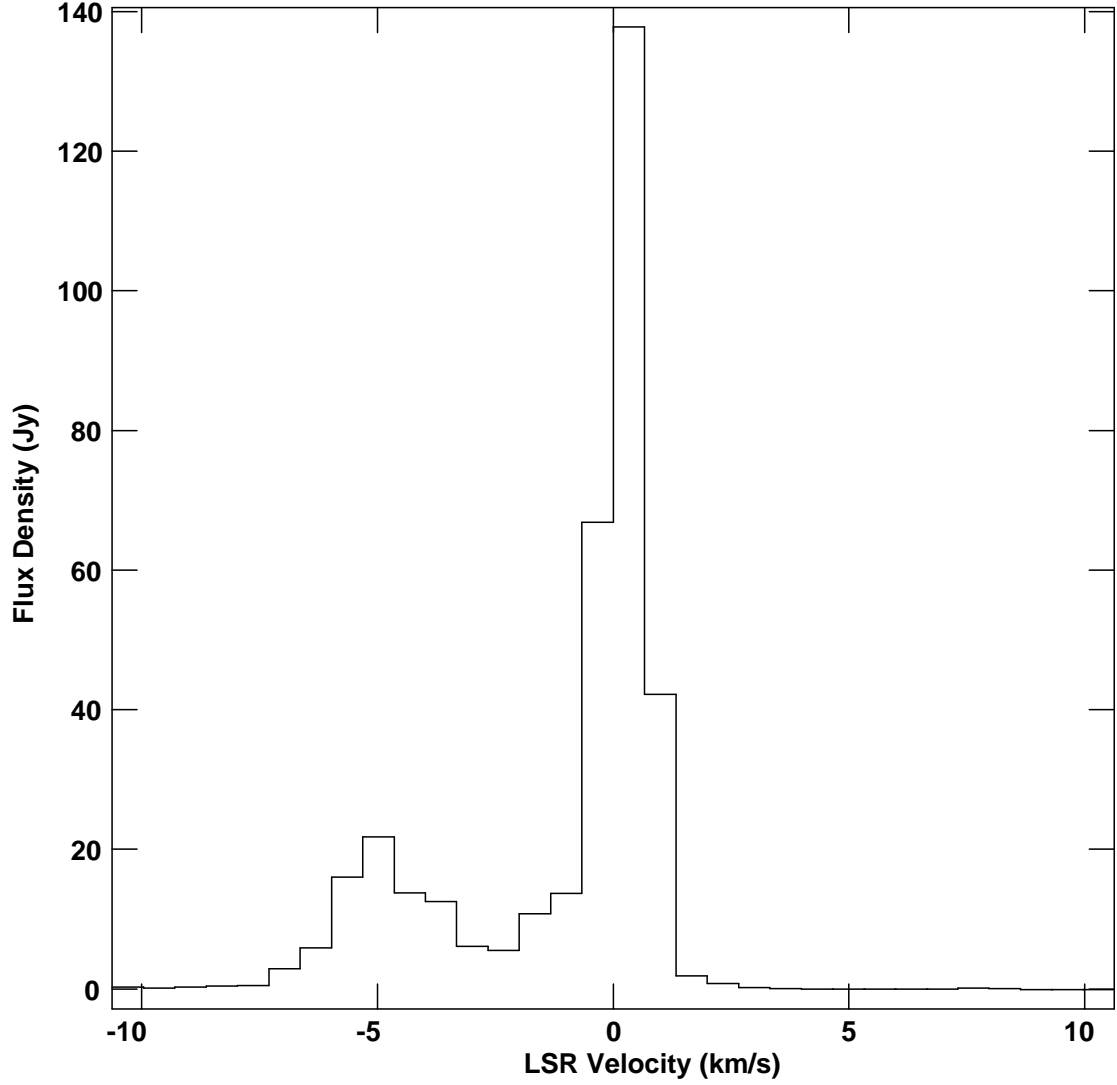} 
\vspace*{16cm}\caption{CH$_3$OH 44$\,$GHz spectrum of the DR21(OH)
region. The spectrum was obtained from a 35\arcsec $\times$ 11\arcsec~
area encompassing the CH$_3$OH maser arcs shown in Figure~4.}
\label{f2}
\end{figure}

\clearpage

\begin{figure}
\includegraphics{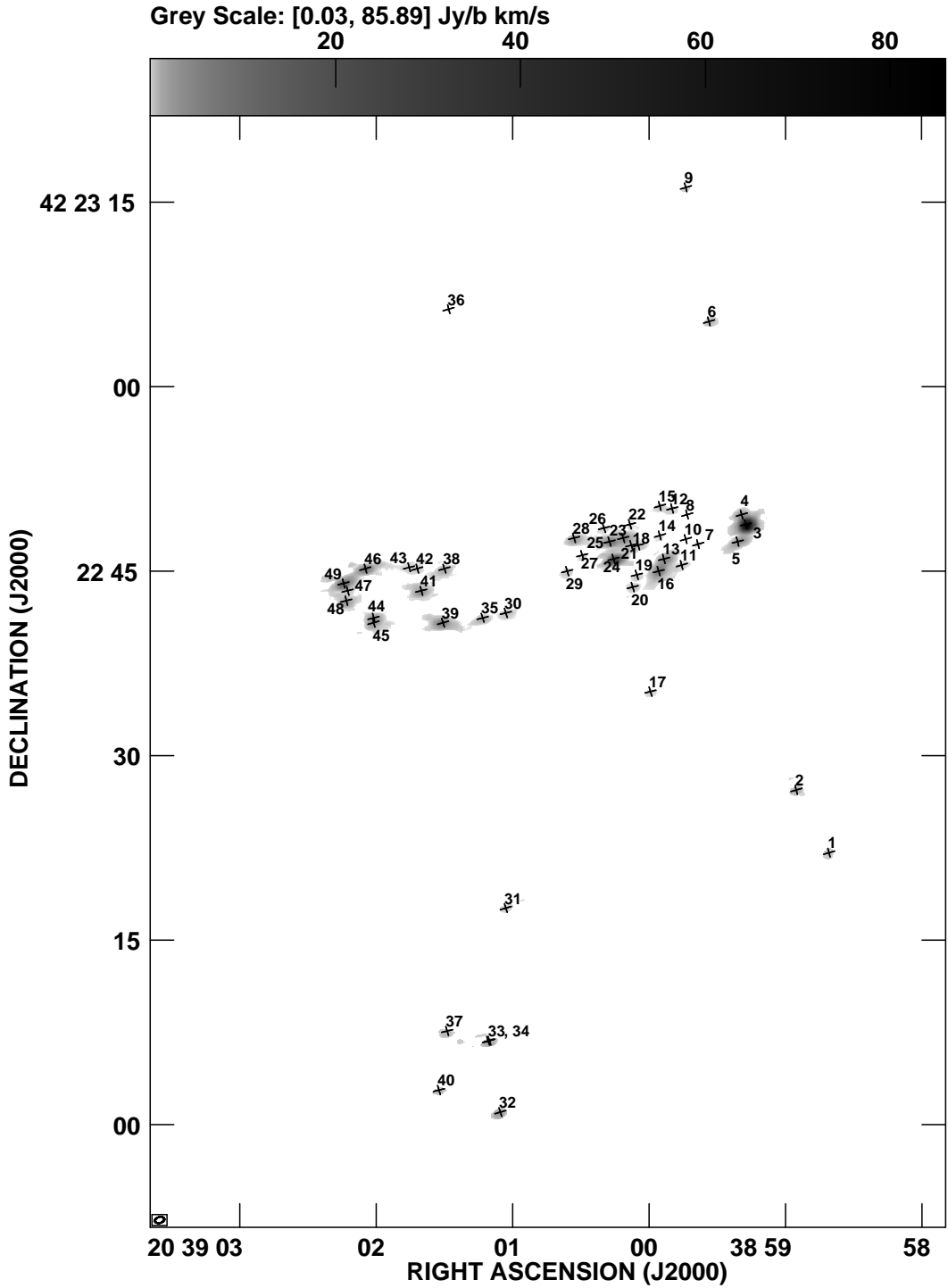} 
\vspace*{18cm}\caption{Integrated intensity (zero velocity moment) of 
the CH$_3$OH 44$\,$GHz masers in the DR21(OH) region. The numbers mark
the position of the masers as given in Table~4.}
\label{f3}
\end{figure}

\clearpage

\begin{figure}
\includegraphics{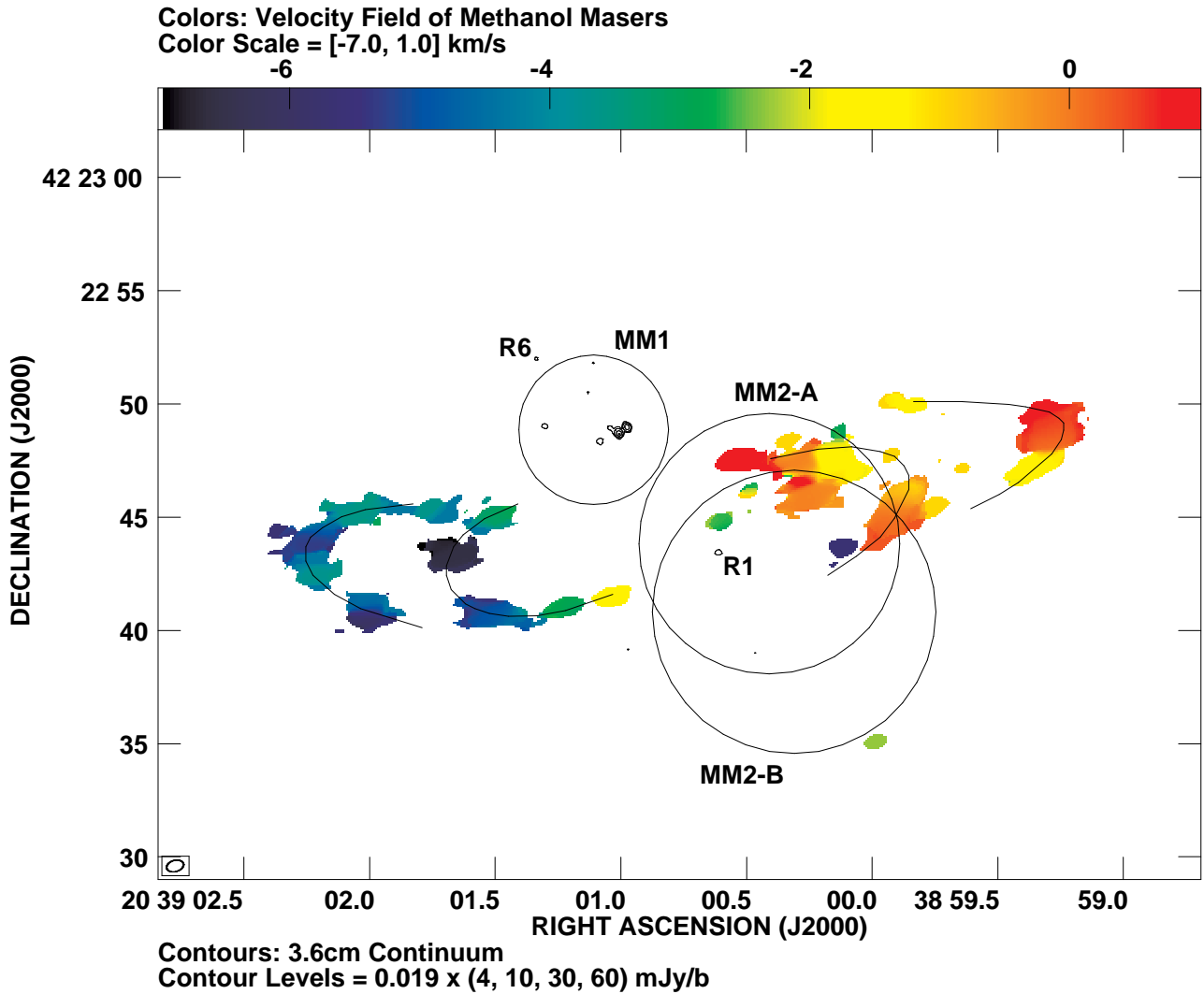} 
\vspace*{18cm}\caption{Velocity field (first velocity moment) of 
the CH$_3$OH 44$\,$GHz masers in the DR21(OH) region. The 3.6$\,$cm continuum
from Figure~1 is shown in contours; sources R1 and R6 are explicitly
marked to facilitate comparison with Figure~1. The location of the 
NH$_3$ molecular cores from Mangum et al. (1992) are shown with circles;
the diameter of the circles equals the major axis of the cores 
as reported by Mangum et al. (1992).}
\label{f4}
\end{figure}

\clearpage

\begin{figure}
\includegraphics{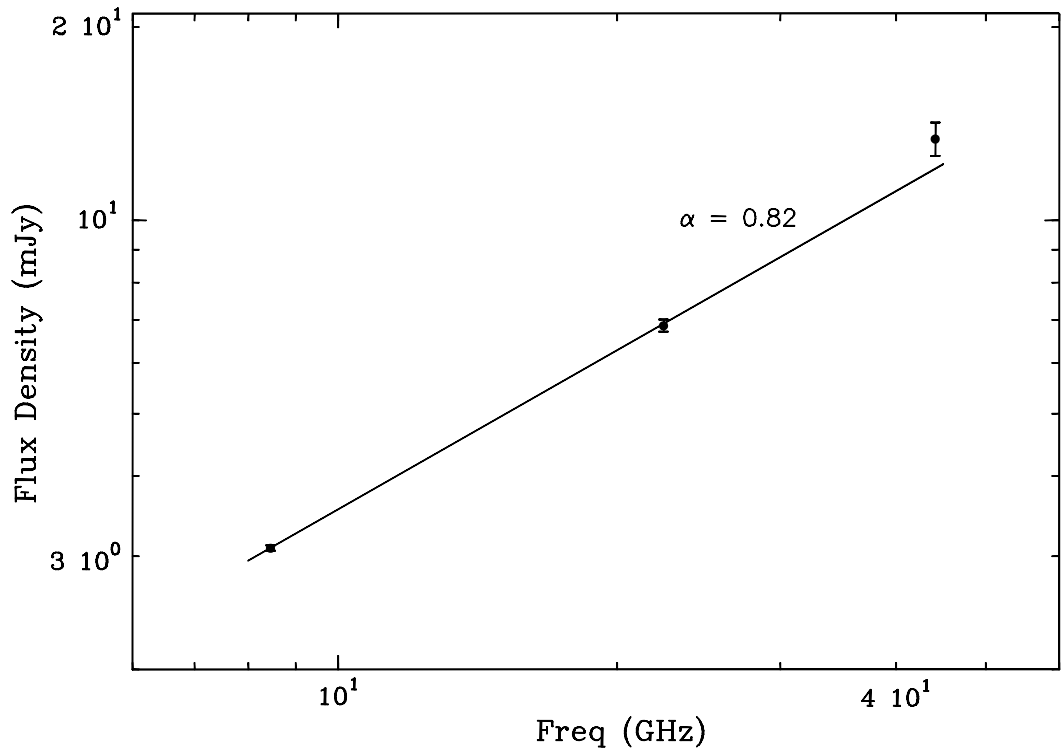} 
\vspace*{11cm}\caption{Radio spectral energy distribution of the MM1 
continuum detected in this work. The flux densities at X and K bands
were obtained after convolving the maps to the synthesized beam
of the Q band observations. A weighted power 
law fit ($S_{\nu}\varpropto\nu^\alpha$)
to the data is shown.
}
\label{f5}
\end{figure}

\clearpage

\begin{figure}
\includegraphics{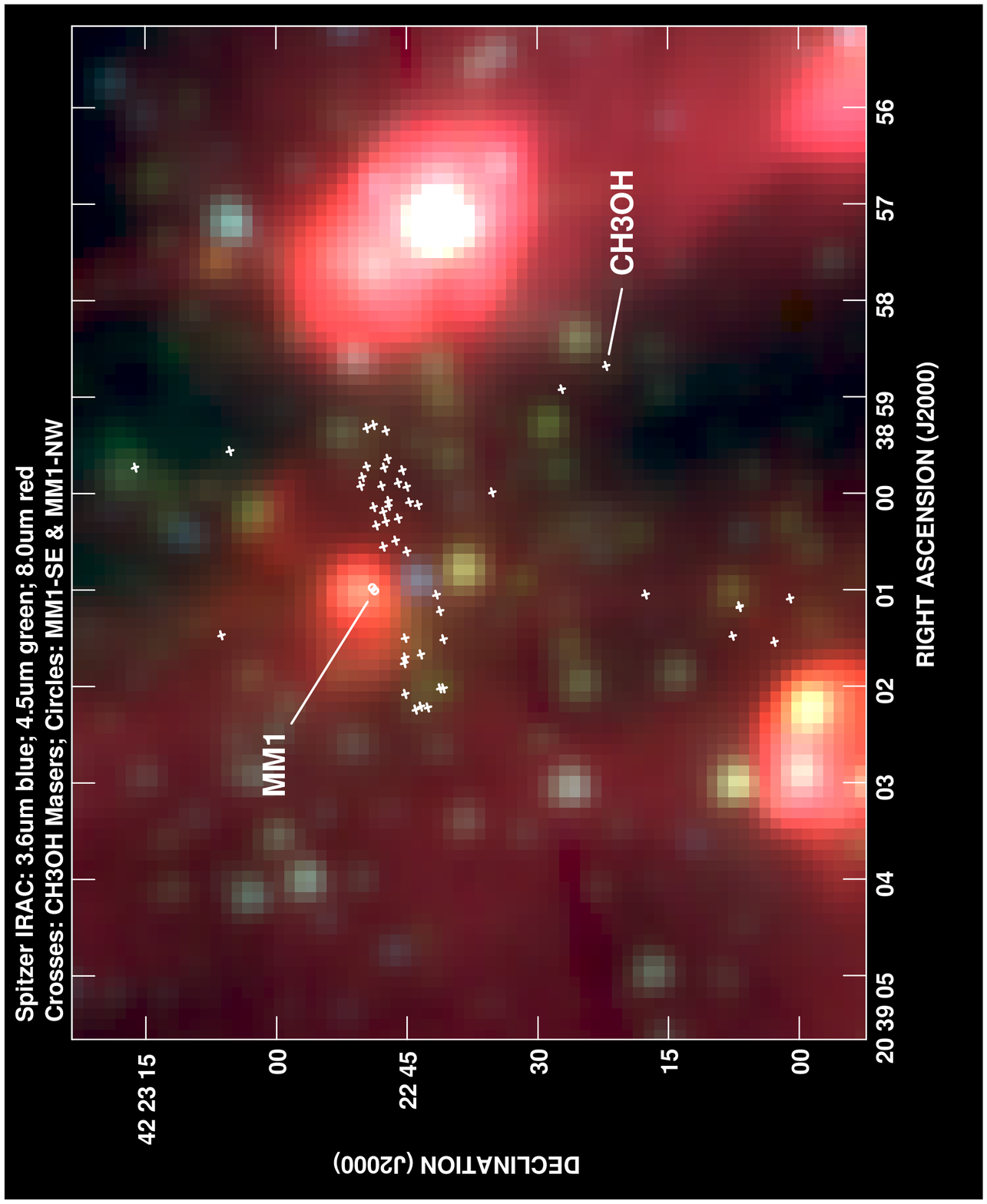} 
\vspace*{11cm}\caption{{\it Spitzer} IRAC image of the DR21(OH) region
(3.6$\,\mu$m blue; 4.5$\,\mu$m green; 8.0$\,\mu$m red). The images 
were retrieved from the {\it Spitzer} archive (http://irsa.ipac.caltech.edu/applications/Spitzer/Spitzer/). The crosses mark the positions of CH$_3$OH
44$\,$GHz masers from this work (Figure~3) and the two circles
show the positions of the radio sources MM1-NW and SE (Figure~1).
Note the 8.0$\mu$m dark area oriented north-south that  
corresponds to the large scale molecular filament in the DR21 region 
(e.g., Kumar et al. 2007).}
\label{f6}
\end{figure}


\begin{thebibliography}{}

\bibitem[]{} Apai, D., Linz, H., Henning, Th., \& 
Stecklum, B. 2005, A\&A, 434, 987

\bibitem[]{} Araya, E., Hofner, P., Kurtz, S., Olmi, L., \& Linz, H.
2008, ApJ, 675, 420

\bibitem[]{} Araya, E., Hofner, P., Olmi, L., Linz, H., Kurtz, S., 
\& Cesaroni, R. 2003, in IAU Symp. 221, Star Formation at High 
Angular Resolution, ed. M. Burton, R. Jayawardhana, \& T. Bourke 
(Cambridge: Cambridge Univ. Press), 489

\bibitem[]{} Arce, H. G., Shepherd, D., Gueth, F., Lee, C.-F., 
Bachiller, R., Rosen, A., \& Beuther, H. 2007, in Protostars \& 
Planets V, ed. B. Reipurth, D. Jewitt, \& K. Keil 
(Tucson: Univ. Arizona Press), 245

\bibitem[]{} Argon, A. L., Reid, M. J., \& Menten, K. M.
2000, ApJS, 129, 159


\bibitem[]{} Bally, J. 2008, in Massive Star Formation: Observations Confront
Theory, ed. H. Beuther, H. Linz, \& Th. Henning (San Francisco: ASPCS), 158


\bibitem[]{887} Beuther, H., Churchwell, E. B., McKee, C. F., \&
Tan, J. C. 2007, in Protostars \& Planets V,
ed. B. Reipurth, D. Jewitt, \& K. Keil
(Tucson: Univ. of Arizona Press), 165


\bibitem[]{} Beuther, H., Schilke, P., Gueth, F., McCaughrean, 
M., Andersen, M., Sridharan, T. K., \& Menten, K. M.
2002, A\&A, 387, 931

\bibitem[]{} Chandler, C. Gear, W. K., \& Chini, R.
1993a, MNRAS, 260, 337

\bibitem[]{} Chandler, C., Moore, T. J. T., Mountain, C. M.,
\& Yamashita, T. 1993b, MNRAS, 261, 694

\bibitem[]{} Cyganowski, C. J., et al. 2008, AJ, 136, 2391

\bibitem[]{} Davis, C. J., Kumar, M. S. N., Sandell, G., 
Froebrich, D. Smith, M. D., \& Currie, M. J. 2007, MNRAS, 374, 29

\bibitem[]{} Devine, D., Bally, J., Reipurth, B., 
Shepherd, D., \& Watson, A. 1999, AJ, 117, 2919

\bibitem[]{} Ducati, J. R., Bevilacqua, C. M., Rembold, S. B., 
\& Ribeiro, D. 2001, ApJ, 558, 309

\bibitem[]{} Eisl\"offel, J., Mundt, R., Ray, T. P., \&
Rodr\'{\i}guez, L. F. 2000, Protostars \& Planets IV, ed. 
V. Mannings, A. Boss \& S. Russell (Tucson: Univ. of Arizona
Press), 815

\bibitem[]{} Ellingsen, S. P. 2006, ApJ, 638, 241

\bibitem[]{} Felli, M., Taylor, G. B., Catarzi, M., Churchwell, E.,
\& Kurtz, S. 1993, A\&ASS, 101, 127

\bibitem[]{} Fish, V. L., Reid, M. J., Argon, A. L., \&
Zheng, X.-W. 2005, ApJS, 160, 220

\bibitem[]{} Garay, G., Mardones, D., Rodr\'{\i}guez, L. F., 
Caselli, P., \& Bourke, T. L. 2002, ApJ, 567, 980

\bibitem[]{} Ghavamian, P., \& Hartigan, P. 1998, ApJ, 501, 687

\bibitem[]{} Girart, J. M., Curiel, S., Rodr\'{\i}guez, L. F.,
\& Cant\'o, J. 2002, RMxA\&A, 38, 169

\bibitem[]{} Harvey-Smith, L., \& Soria-Ruiz, R. 2008, MNRAS, 391, 1273

\bibitem[]{} Harvey-Smith, L., Soria-Ruiz, R., Duarte-Cabral, A.,
\& Cohen, R. J. 2008, MNRAS, 384, 719

\bibitem[]{} Hillenbrand, L. A.,  et al. 1998, AJ, 116, 1816

\bibitem[]{} Hofner, P., Cesaroni, R., Olmi, L., Rodr\'{\i}guez, L. F.,
Mart\'{\i}, J., \& Araya, E. 2007, A\&A, 465, 197

\bibitem[]{} Ignace, R., \& Churchwell, E., 2004, ApJ, 610, 351

\bibitem[]{} Johnston, K. J., Henkel, C. \& Wilson, T. L. 
1984, ApJ, 285, L85

\bibitem[]{} Kogan, L., \& Slysh, V. 1998, ApJ, 497, 800

\bibitem[]{} Kumar, M. S. N., Davis, C. J., Grave, J. M. C., Ferreira, B.
\& Froebrich, D. 2007, MNRAS, 374, 54

\bibitem[]{} Kurtz, S., Hofner, P., \& Vargas-\'Alvarez, C. 
2004, ApJS, 155, 149

\bibitem[]{} Lai, S.-P., Girart, J. M., \& Crutcher, R. M.
2003, ApJ, 598, 392

\bibitem[]{} Liechti, S., \& Walmsley, C. M. 1997, A\&A, 321, 625

\bibitem[]{} Mangum, J. G., Wootten, A., \& Mundy, L. G. 1992, ApJ, 388, 467 

\bibitem[]{} Mangum, J. G., Wootten, A., \& Mundy, L. G. 1991, 
ApJ, 378, 576

\bibitem[]{} Marston, A. P., et al. 2004, ApJS, 154, 333

\bibitem[]{} Mart\'{\i}, J., Rodr\'{\i}guez, L. F., \& Reipurth, B. 
1995, ApJ, 449, 184

\bibitem[]{} Mart\'{\i}, J., Rodr\'{\i}guez, L. F., \& Reipurth, B.
1998, ApJ, 502, 337

\bibitem[]{} Mathis, J. S. 1990, ARA\&A, 28, 37

\bibitem[]{} Meyer, M. R., Calvet, N., \& 
Hillenbrand, L. A. 1997, AJ, 114, 288

\bibitem[]{} Moscadelli, L., Cesaroni, R., \& Rioja, M. J. 2005, A\&A, 
438, 889

\bibitem[]{} Motte, F., Bontemps, S., Schilke, P., Lis, D. C., 
Schneider, N., \& Menten, K. M. 2005, in Massive Star 
Birth: A Crossroads of Astrophysics, IAU Symp 227, ed. R. Cesaroni, 
M. Felli, E. Churchwell, \& C. M. Walmsley
(Cambridge: Cambridge Univ. Press), 151

\bibitem[]{} Motte, F., Bontemps, S., Schilke, P., Schneider, N., Menten, K. M., 
\& Brogui\`ere, D. 2007, A\&A, 476, 1243

\bibitem[]{} Odenwald, S. F., \& Schwartz, P. R. 1993, ApJ, 405, 706

\bibitem[]{} Plambeck, R. L. \& Menten, K. M. 1990, ApJ, 364, 555

\bibitem[]{} Pratap, P., Shute, P. A., Keane, Th. C., 
Battersby, C., \& Sterling, S. 2008, AJ, 135, 1718

\bibitem[]{} Reipurth, B., Rodr\'{\i}guez, L. F., Anglada, G., 
\& Bally, J. 2002, AJ, 124, 1045

\bibitem[]{} Richardson, K. J., Sandell, G., Cunningham, C. T., \&
Davies, S. R. 1994, A\&A, 286, 555

\bibitem[]{} Smith, H. A., Hora, J. L., Marengo, M., \& Pipher, J. L.
2006, ApJ, 645, 1264

\bibitem[]{} Vall\'ee, J. P., \& Fiege, J. D. 2006, ApJ, 636, 332

\bibitem[]{} Zapata, L. A., Rodr\'{\i}guez, L. F., 
Kurtz, S. E., O'Dell, C. R., \& Ho, P. T. P. 2004, ApJ, 610, L121

\end{thebibliography}
\end{document}